\documentclass[aps,pra,showpacs,twocolumn]{revtex4-1}
\usepackage{graphicx,psfrag,amsmath,amssymb,amsfonts,latexsym,color,epsf,graphpap,dcolumn}

\let\oldtabular\tabular 
\renewcommand{\tabular}{\footnotesize\oldtabular}

\begin{document}
\title{Casimir Free Energy at High Temperatures: Grounded vs Isolated
Conductors}
\author{C.~D.~Fosco$^{a}$, F.~C.~Lombardo$^{b}$ and F.~D.~Mazzitelli$^{a}$}
\affiliation{$^a$Centro At\'omico Bariloche and Instituto Balseiro, CONICET,
Comisi\'on Nacional de Energ\'\i a At\'omica, R8402AGP Bariloche, Argentina.\\
$^b$Departamento de F\'\i sica {\it Juan Jos\'e
Giambiagi}, FCEyN UBA and IFIBA CONICET-UBA, Facultad de Ciencias Exactas y Naturales,
Ciudad Universitaria, Pabell\' on I, 1428 Buenos Aires, Argentina}
\date{today}
\begin{abstract}
\noindent 
We evaluate the difference between the Casimir free energies corresponding
to either grounded or isolated perfect conductors, at high temperatures.
We show that a general and simple expression for that difference can be
given, in terms of the electrostatic capacitance matrix for the system of
conductors.
For the case of close conductors, we provide approximate expressions for
that difference, by evaluating the capacitance matrix using the proximity
force approximation.  

Since the high-temperature limit for the Casimir free energy for a medium
described by a frequency-dependent conductivity diverging at zero frequency
coincides with that of an isolated conductor, our results may shed light on
the corrections to the Casimir force in the presence of real materials.  
\end{abstract} 
\maketitle
\section{Introduction}\label{sec:intro}
Casimir forces and related phenomena constitute remarkable macroscopic
manifestation of zero point or thermal fluctuations of the electromagnetic
field. Different high precision experiments have been implemented in recent
years in order to measure the Casimir force with ever increasing
detail~\cite{books}. 
In spite of these efforts, the corresponding comparison between theory and
experiment has not yet been, however, entirely satisfying~\cite{Mostep2015}. This
suggests that further theoretical and experimental developments may be
required in order to tackle some of the long standing puzzles 
which arise in realistic descriptions of the
forces and their detailed properties.

In this paper, we find a general expression for the difference, $\Delta F$,
between the high temperature free energies for two different cases,
according to whether the conductors are: a) grounded or b) isolated.
Albeit this is a question which has been partially addressed in previous
works \cite{Golestanian2000, BimonteEmig2012}, we want to present a fuller
answer here, allowing us to consider different concrete
examples.

As we shall see, isolated perfect conductors can be used to describe, in
the high temperature limit, real materials with a permittivity diverging in
the zero-frequency limit.  Therefore, $\Delta F$ may be used to account, in
those cases, for real material corrections to the Casimir effect of
grounded conductors at high temperatures.  Thus, even though most
experimental setups involve grounded conductors (in order to minimize
spurious electrostatic effects), the question
we address may be relevant to account for those corrections, apart from its conceptual interest. 

\section{Free energy for the electromagnetic field}\label{free}
Since our focus shall be on the high temperature limit of the free energy,
we begin by deriving the expression for the free energy $F$ for the quantum
electromagnetic (EM) field at a finite temperature.  It is  a function of
the inverse temperature $\beta = T^{-1}$ (in our conventions, Boltzmann's
constant $k_B \equiv 1$).  $F$  may be written in terms of the partition
function, ${\mathcal Z}$ as follows:
\begin{equation}
	F\,=\,-\frac{1}{\beta}\,\log\big[\frac{{\mathcal
	Z}}{{\mathcal Z}_0}\big]\;,
\end{equation}
where the denominator, ${\mathcal Z}_0$, denotes the partition
function for the free (i.e., in the absence of media) EM field. 
The effect of that denominator is to subtract the free energy of a free Bose gas of
photons in the absence of the mirrors, which does not contribute to the
force between them.  

In the Matsubara formalism, a functional integral expression
for the partition function ${\mathcal Z}$ can be constructed by
integrating over  field configurations depending on the spatial
coordinates ${\mathbf x}$ and the imaginary time $x_0 \equiv \tau$. The
fields are periodic, with period $\beta$, in the imaginary time. Denoting
by $A=(A_\mu)$, ($\mu=0,1,2,3$) the $4$-potential in Euclidean (imaginary
time) spacetime, ${\mathcal Z}$ is given by:
\begin{equation}\label{eq:defzbeta}
{\mathcal Z} \;=\; \int \big[{\mathcal D}A\big] \;
e^{-{\mathcal S}_{\rm inv}(A)}
\end{equation} 
where ${\mathcal S}_{\rm inv}(A)$ is the gauge-invariant action for $A$, 
while $\big[{\mathcal D}A\big]$ is used to denote the functional
integration measure including gauge fixing.
 
In terms of the components of the field strength tensor
$F_{\mu\nu}=\partial_\mu A_\nu - \partial_\nu A_\mu$, the form of the
gauge-invariant action in the presence of real materials is:
\begin{eqnarray}
{\mathcal S}_{\rm inv}(A) &=&\int_0^\beta d\tau \int_0^\beta d\tau' \int d^3{\mathbf x} 
\big[  \frac{1}{2} F_{0j}(\tau,{\mathbf x})
\epsilon(\tau-\tau', {\mathbf x}) \label{eq:defsinv} \\ &\times & F_{0j}(\tau',{\mathbf x}) 
 +  \frac{1}{4} F_{ij}(\tau,{\mathbf x})
\mu^{-1}(\tau-\tau', {\mathbf x})  F_{ij} (\tau,{\mathbf x}) \big] \;,\nonumber
\end{eqnarray}
where indices from the middle of the Roman alphabet run over spatial
indices (Einstein summation convention has been adopted), and
$\epsilon(\tau-\tau' , {\mathbf x})$ and $\mu(\tau-\tau' , {\mathbf x})$ 
denote the Euclidean versions of the permittivity and permeability,
respectively ($\mu^{-1}$ is the inverse integral kernel of $\mu$, with
respect to its time-like arguments). 
Space locality of those response functions has been assumed implicitly.

It is rather useful to adopt mixed Fourier transformations for the fields, as well
as for the response functions:
\begin{eqnarray}\label{eq:fou1}
A_\mu (\tau,{\mathbf x}) \;&=&\; \frac{1}{\beta} \,
\sum_{n=-\infty}^{+\infty} \widetilde{A}_\mu^{(n)}({\mathbf x}) \, e^{i \omega_n
\tau} \nonumber\\
\epsilon(\tau-\tau',{\mathbf x}) \;&=&\; \frac{1}{\beta} \,
\sum_{n=-\infty}^{+\infty} \widetilde{\epsilon}^{(n)}({\mathbf x}) \, e^{i \omega_n
(\tau-\tau')} \nonumber\\
\mu(\tau-\tau',{\mathbf x}) \;&=&\; \frac{1}{\beta} \,
\sum_{n=-\infty}^{+\infty} \widetilde{\mu}^{(n)}({\mathbf x}) \, e^{i \omega_n
(\tau-\tau')} 
\end{eqnarray}   
where $\omega_n \equiv \frac{2\pi n}{\beta}$ ($n \in {\mathbb Z}$) are the
Matsubara frequencies. Note that, with the convention above for the
definition of the Fourier expansion, $\widetilde{A}_\mu^{(n)}$ is
a dimensionless field.

The high temperature (classical) limit is dominated, for a Bose field, 
by the $n=0$ Matsubara mode.  In a previous work \cite{Foscoetal2015}, we
have shown that the zero mode free energy can be written as
\begin{equation}
F=F_s +F_v\, ,
\end{equation} 
where $F_s$ corresponds to the free energy of a scalar field in $2+1$
(Euclidean) dimensions
\begin{equation}
e^{- \beta \, F_s} \,=\, \int {\mathcal D} \widetilde{A}_0^{(0)} \,
e^{- \frac{1}{2 \beta}
\int d^3{\mathbf x} \, \widetilde{\epsilon}^{(0)}({\mathbf x}) 
(\partial_j \widetilde{A}_0^{(0)})^2 }\, ,
\end{equation}
while $F_v$ is the free energy of a vector field in $2+1$ dimensions 
\begin{equation}
e^{- \beta F_v(\psi)} = \int {\mathcal D}
\widetilde{A}_j^{(0)} 
e^{- \frac{1}{\beta} \int d^3{\mathbf x} [ \frac{1}{4
\widetilde{\mu}^{(0)}({\mathbf x})} (\widetilde{F}_{jk}^{(0)})^2 
+ \frac{1}{2}  \Omega_0^2({\mathbf x}) (\widetilde{A}_j^{(0)})^2]}.
\end{equation}
Here, we have introduced the object:
\begin{equation}
\Omega_0^2({\mathbf x}) \,\equiv\, \lim_{n\to 0} \,
\big[ \omega_n^2 \,\widetilde{\epsilon}^{(n)}({\mathbf x}) \big]
\end{equation}
(note that $\widetilde{\epsilon}^{(0)}$, $\widetilde{\mu}^{(0)}$ and
$\Omega_0$ are model-dependent). 

Let us now discuss the limit of perfectly-conducting materials, from the
point of view of the scalar and vector contributions: regarding the field
$\widetilde{A}^{(0)}_0$, which behaves as a $2+1$ dimensional scalar,  the
infinite permittivity limit implies that  its gradient inside the regions
occupied by the material bodies vanishes identically. Therefore, the field
is constant in those regions. On the other hand, if the conductors are grounded,
those constants must vanish, so that the field itself is zero.  Namely, the
scalar field is subjected to Dirichlet boundary conditions, corresponding
to the transverse magnetic (TM) EM mode.  If the conductors are isolated,
the field can take any value in each non-vacuum region. In this case, the
functional integral should be performed over all possible configurations,
including arbitrary (constant) values on the surfaces of
the conducting bodies.

The vector zero mode, on the other hand, behaves as an EM field in $2+1$
dimensions. If $\Omega_0$ tends to infinity, then the   EM field will
vanish identically on the regions filled up by media. It then satisfies
perfect conductor boundary conditions. We have shown this to be equivalent
to a real scalar field with Neumann conditions~\cite{Neumann2+1},
corresponding to the transverse electric (TE) EM mode. 

There is a well known subtle point in the case of real materials, which
manifests itself when considering two typical models for the permittivity,
namely, the Drude or plasma models, where the permittivity diverges in the zero frequency
limit. Therefore, in both cases the associated TM mode contribution is tantamount to that
of a scalar field in the presence of an isolated perfect conductor. There
is a difference, however, in the TE mode contribution for both models.
Indeed, since $\Omega_0$ vanishes in the Drude model, there is no TE
contribution to the Casimir free energy, whilst the plasma model generates a
non vanishing TE mode. The latter coincides with that of a perfect conductor
in the limit of a high plasma frequency.
 
\section{Grounded vs isolated free energies}
In what follows
we will consider in detail the scalar TM contribution, aiming to obtain the
difference
\begin{equation}
\Delta F = F_s^{(g)}-F_s^{(i)}
\end{equation}
between grounded and isolated perfect conductor boundary conditions. 
In view of the discussion at the end of the previous Section, $\Delta F$ describes
the difference between the scalar field term in the free energy of a system of 
grounded perfect conductors and that corresponding to the same geometry but
involving materials which are described by  Drude or plasma models. 

In order to simplify the notation, we adopt a simpler notation for the only
field we have do deal with henceforward, namely:
$\widetilde{A}^{(0)}_0\equiv\phi$ (we recall that $\phi$ is dimensionless,
because of the definition for the Fourier transforms used in
Eq.(\ref{eq:fou1})).  Regarding the geometry, we assume that the system
under consideration consists of $N$ conductors, each one occupying a
volume $V_\alpha$ enclosed by a surface
$S_\alpha$, with $\alpha=1,2, ....N.$

An intermediate object that may be conveniently used as an ingredient to obtain both the
grounded and isolated conductors partition functions, is a
partition function where the (constant) value of $\phi$ on each surface $S_\alpha$ is
fixed to a given but otherwise arbitrary value $\phi_\alpha$. The partition function for these particular
boundary conditions is denoted by,
\begin{equation}
 {\mathcal Z}[\{\phi_\alpha\}]=\, \int {\mathcal D} \phi \,
e^{- \frac{1}{2 \beta}
\int d^3{\mathbf x} \, 
(\partial_j\phi )^2 }\prod_{\alpha=1}^N\delta[\phi\vert_{S_\alpha}-\phi_\alpha] \; .
\end{equation}

Thus, we may obtain the partition functions corresponding to  grounded
(${\mathcal Z}^{(g)}$) and isolated (${\mathcal Z}^{(i)}$) conductors as follows \cite{Golestanian2000}:
\begin{equation}
{\mathcal Z}^{(g)}\;=\;{\mathcal Z}[\{\phi_\alpha \}]\Big|_{\phi_\alpha =
0}
\end{equation}
and
\begin{equation} \label{iso}
{\mathcal
Z}^{(i)}\;=\;\int_{-\infty}^{\infty}\left(\prod_{\alpha=1}^{N}d\phi_\alpha\right)
\; {\mathcal Z}[\{\phi_\alpha \}]\;.
\end{equation}

It is self-evident that Eq.(\ref{iso}) does not correspond to grounded
conductors, since the values of the potentials at each
surface are not fixed to zero; rather they have to be integrated out.  
One can show explicitly that the result of that integration corresponds to
a situation in which the total charge of each conductor is zero, with
vanishing charge fluctuations (there are of course $\phi_\alpha$
fluctuations). We present a derivation of this property, within the context
of our approach, in the Appendix (see also~\cite{Golestanian2000b}).

In view of its relevance to both the grounded and isolated limits, let us
then compute ${\mathcal Z}[\{\phi_\alpha \}]$. To that end, it is
convenient to perform a shift (translation) in the integration variables:
$\phi ({\mathbf x})=\tilde\phi ({\mathbf x}) +\varphi({\mathbf x})$,
where $\tilde\phi$ is the (unique) solution of the classical electrostatic problem
with  prescribed boundary conditions for the potential on the
conductors:
\begin{equation}\label{shift}
\nabla^2\tilde\phi({\mathbf x})=0\,\, \, ,\,\,  \tilde\phi\vert_{S_\alpha}=\phi_\alpha\,\, \, ,\,
\end{equation}
and $\varphi({\mathbf x})$ is a scalar field satisfying Dirichlet boundary
conditions. It is rather straightforward to show that, after the shift, we
have:
\begin{eqnarray}
 {\mathcal Z}[\{\phi_\alpha
\}]&=&e^{-\frac{1}{2\beta}\sum_{\gamma\delta}C_{\gamma\delta}\phi_\gamma\phi_\delta}\,
 \int {\mathcal D} \varphi \, e^{- \frac{1}{2 \beta}
\int d^3{\mathbf x} 
(\partial_j\varphi )^2 }\nonumber \\
& \times &\prod_{\alpha=1}^N\delta[\varphi\vert_{S_\alpha}]
=
e^{-\frac{1}{2\beta}\sum_{\gamma\delta}C_{\gamma\delta}\phi_\gamma\phi_\delta}
\, {\mathcal Z}^{(g)}\; ,
\end{eqnarray}
where the $C_{\gamma\delta}$ denote the capacitance coefficients of the
system of conductors. 

As discussed previously, ${\mathcal
Z}^{(i)}$ is obtained by performing a  Gaussian integral over the constant
values of the potential on the conductors, obtaining
\begin{equation}
\Delta F=-\frac{1}{2\beta} \log[\det ({\mathbb C})/\beta^N]\;,\label{DeltaF}
\end{equation}
where $\mathbb C$ is the capacitance matrix. This is the main result of this work.
In what follows we will omit the factor $\beta^N$ inside the logarithm,
since it is irrelevant when computing the Casimir forces between conductors. 

\section{Examples and PFA approximation}

In this Section we evaluate $\Delta F$ for some particular geometries, and
analyze its behavior at long and short distances.  The latter is
elucidated by using an estimation of the capacitance matrix, obtained by using the proximity
force approximation (PFA) \cite{Derjaguin}.

\subsection{Sphere-sphere geometry}

Let us consider two facing spheres of equal radius $a$, separated by a
distance $d$ between centers. The elements of the capacitance matrix for
this geometry are
given by \cite{Smythe}
\begin{eqnarray} C_{11} = C_{22} &=& a \sinh\psi \sum_{n=1}^{\infty} {\rm
csch}\left((2 n - 1)\beta\right), \nonumber \\ 
C_{12} = C_{21} &=& -  a \sinh\psi \sum_{n=1}^{\infty} {\rm  csch}\left(2
n\psi\right),\label{exact1}\end{eqnarray}
where $\cosh\psi \equiv d/2a$. Inserting Eq.(\ref{exact1}) into
Eq.(\ref{DeltaF}) one obtains an exact analytic expression for
$\Delta F$ in this geometry.

When both spheres are very close, $d\rightarrow 2 a$, we  define 
$\xi = (d - 2a)/2a$ and take the limit $\xi \rightarrow 0$ in Eqs.
(\ref{exact1}). It can be shown that \cite{electros} 
\begin{eqnarray} 
C_{11} = C_{22} &=&  a \left(-\frac{1}{4} \log\xi + \frac{\gamma}{2} + \frac{3}{4} \log 2 + {\cal O}(\xi \log\xi) \right) 
, \nonumber \\ C_{12} = C_{21} &=&   a \left(\frac{1}{4} \log\xi - \frac{\gamma}{2} + \frac{1}{4} \log 2 + 
{\cal O}(\xi \log\xi) \right)\end{eqnarray}
where $\gamma$ denotes the Euler-Mascheroni constant. In this approximation:
\begin{equation}
\beta \Delta F \approx -\frac{1}{2} \log\left(- \log\xi\right).\label{loglog}
\end{equation}

We note that the last result coincides with  the one obtained in
\cite{BimonteEmig2012},  where the authors computed the high temperature
Casimir free energies for the same geometry, considering Dirichlet and metallic
boundary conditions, the latter described by a Drude model. It has also
been shown there that, in the short distance limit, one has  
\begin{equation}
\beta F_s^{(g)}\approx -\frac{\zeta(3)}{16\, \xi}+\frac{1}{48}\log\xi \, ,
\end{equation}
where the leading term is the usual PFA, while the next to leading order
(NTLO) can be obtained using the derivative expansion approach
\cite{denos,deothers}.  We see that the difference $\Delta F$ is much
smaller than the NTLO as $\xi\to 0$.  Note, however,  that due to the
presence of the double logarithm this will only happen for exceedingly small values of $\xi$
(and therefore the double logarithmic term becomes the main correction to the PFA
for typical values of $d$ and  $a$).

Let us now consider the opposite limit,  $d \gg a$ (large separation),
where the capacitance coefficients have the expansions:
\begin{eqnarray} 
C_{11} = C_{22} &=&  a \left(1 + \frac{a^2}{d^2} + \frac{2 a^4}{d^4} + ... \right) 
, \nonumber \\ C_{12} = C_{21} &=& -  a \left(\frac{a}{d} + \frac{a^3}{d^3} + \frac{3 a^5}{d^5} +... \right)  .\end{eqnarray}
 Inserting this result into Eq.(\ref{DeltaF}) we obtain:
\begin{equation}
\beta \Delta F  \approx -\frac{1}{2}   \left( \frac{a^2}{d^2} + \frac{5}{2} \frac{a^4}{d^4} \right).\end{equation}

The free energy for grounded spheres has been obtained in
Ref.\cite{BimonteEmig2012}. Performing an expansion of their exact result
in
the large distance limit, we get $F^{(g)}_s \approx \Delta F$, 
and therefore $ \ F^{(i)}_s = {\cal O}\left( \left(\frac{a}{d}\right)^6\right)$, 
which shows that the interaction between isolated spheres is dominated by
the dipole-dipole interaction, as discussed in Ref.\cite{Golestanian2000}
for compact objects using a  multipole expansion. 

This example illustrates a general characteristic of the difference between
the free energies for grounded and isolated objects.  While at short
distances both $F_s^{(g)}$ and $F_s^{(i)}$ have the same leading order
behavior, at long distances the fluctuations of the charges (that occur
for  grounded conductors and not for isolated ones), radically change the
nature of the leading interaction between conductors.

\subsection{Sphere-plane geometry}

We shall now consider a sphere of radius $a$, whose center is at a distance
$d$ from an infinite plane. The elements of the 
capacitance matrix can be obtained as a limiting case of a geometry involving  two 
separated spheres with different radii $a$ and $b$, in the limit
$b\rightarrow\infty$ \cite{Smythe}. In this situation,  $C_{22}\approx b$
while 
\begin{equation}
C_{11} =  -C_{12} = a \sinh\alpha \sum_{n= 1}^{\infty} {\rm csch}(n \alpha)
\;, 
\end{equation}
where $ \cosh\alpha = d/a$.  Therefore, the difference between free energies reads
\begin{equation}
\beta \Delta F = -\frac{1}{2} \log (C_{11}b-C_{11}^2)= -\frac{1}{2} \log C_{11} + {\rm const}
\end{equation}
where the constant, independent of $d$ and diverging as
$b\rightarrow\infty$, is irrelevant  for the computation of the force
between sphere and plane.

In the short distance limit $\alpha\rightarrow 0$,  the sum that defines
the coefficient $C_{11}$ can be approximated by
\begin{equation}
  \sum_{n= 1}^{\infty} {\rm csch}(n \alpha) \approx  \int_{1}^{\infty} dn\;  {\rm csch}(n \alpha) = \frac{1}{\alpha}\log[{\rm coth}(\frac{\alpha}{2})]\,
  \end{equation}
and then
\begin{equation}
\beta \Delta F \approx -\frac{1}{2} \log[\frac{1}{2}\log(\frac{a}{d-a})]\, .
\end{equation}
Once again, there is a double log term in the free energy for
isolated objects. This behavior has been found numerically for the same  geometry,
when considering the classical limit of the Casimir interaction for Drude
metallic boundary conditions \cite{Paulo}.  
\subsection{General case: two close conductors}

In view of the examples above, the question presents itself about  whether
the appearance of double logarithms in $\Delta F$ at close distances is
a general feature or not, i.e. if they always appear (independently of the
geometry of the conductors involved).
Exploration of other simple geometries shows that this is not the case.
Indeed, the determinant of $\mathbb C$ does not  show a logarithmic
behavior at short distances for some elementary examples, like concentric cylinders
or concentric spheres.  One can show that this is also the case for
eccentric cylinders or spheres, as well as for a cylinder in front of a plane.

In a general case, assuming that the geometry is such that one can use the
PFA  to estimate the
capacitance matrix elements, the
electrostatic energy between two conductors held at potentials $\phi_1$ and
$\phi_2$ respectively can be approximated by \cite{annphys} 
\begin{equation}
U_{PFA}=\frac{1}{2}(\phi_1-\phi_2)^2\int d^2{\mathbf x}\; 
\frac{1}{d({\mathbf x})}\equiv \frac{1}{2}C_{\rm PFA}(\phi_1-\phi_2)^2\, ,
\end{equation}
where $d({\mathbf x})$ denotes the local distance between facing surface
elements on both conductors. Note that, in this approximation, we have
$C_{11} =C_{22}=-C_{12}= -C_{21}=C_{\rm PFA} $ and therefore $\det{\mathbb
C}$ vanishes.
 
In general, including departures from the PFA result, we will have $C_{11}
= C_{\rm PFA} + \Delta_{11}$, $C_{22} =
C_{\rm PFA} + \Delta_{22}$, and $C_{12} = -   C_{\rm PFA} + \Delta_{12}$, 
with $\Delta_{\alpha\beta}$ denoting the contributions 
coming from the subleading corrections.  Therefore
\begin{eqnarray}\label{FPFA}
\beta \Delta F & \approx &-\frac{1}{2}\log (C_{\rm PFA})  
-\frac{1}{2}\log (\Delta_{11}+\Delta_{22}-2\Delta_{12})\nonumber \\ &\approx &-\frac{1}{2}\log (C_{\rm PFA})\;,
\end{eqnarray}
where we have assumed that the contributions coming from the subleading
corrections are much smaller than the leading PFA term \cite{foot}. From
Eq.(\ref{FPFA}) we can derive the form of the short distance behavior, by
using the corresponding expressions for $C_{PFA}$. They are shown, for all
the examples mentioned above, in Table I.
\begin{table}
 \begin{center}
\begin{tabular}{  | l | l |}
    \hline
    Geometry & $C_{\rm PFA}$   \\ \hline
    Sphere - sphere & \,$-\frac{a}{4}\log\frac{h}{a}$  \\ \hline
    Sphere - plane &$\, -\frac{a}{2}\log\frac{h}{a}$   \\ \hline
     Concentric spheres & \,\, \,\,\,\, $\frac{a^2}{h}$  \\ \hline 
     Concentric cylinders & $\,\,\, \,\,\,\,\,\,\frac{La}{2h}$ \\ \hline      
      Cylinder - plane & $\,\,\frac{L}{4\sqrt 2}\sqrt{\frac{a}{h}}$ \\ \hline
      Eccentric cylinders & $\frac{L}{2}\sqrt{\frac{ab}{2(b-a)h}}$ \\ \hline
       \end{tabular}
\end{center}
\caption{Capacitance matrix elements in the PFA for different geometries 
$C_{\rm PFA}=C_{11}=C_{22}=-C_{12}$. $a$ and $b$ denote the radii of spheres or cylinders, $h$ is the distance between conducting surfaces, and $L$ is the cylinders' length.}
\end{table}

\section{Discussion}\label{sec:conc}
In this paper we have computed the difference between the high temperature
Casimir free energies for a system of conductors, when these
are either grounded or isolated. We have shown that that difference comes
from the TM Matsubara zero mode of the electromagnetic field, which can be
described by a single scalar field. When the conductors are grounded, the
scalar field satisfies Dirichlet boundary conditions. On the other hand,
when the conductors are isolated, the scalar field may take any constant
value on the surface of each conductor, and those constant values have to
be integrated.
Precisely because of that constant-potential integration, the difference
$\Delta F$ becomes proportional to $\log\det{\mathbb C}$, where
$\mathbb C$ denotes the (electrostatic) capacitance matrix of the system.  

We have evaluated explicitly $\Delta F$ for particular geometries, and found a
general expression for the case of two close conductors, using the PFA.
Note that the use of the PFA for the approximate evaluation of $\Delta F$
could be convenient to derive, for example, the free energy for isolated
conductors based on the knowledge of the result corresponding to grounded
ones. The latter could be known by the use of any other method, not
necessarily the PFA.  

Essentially the same problem of evaluating the difference between the two
free energies we have considered has been studied before
\cite{Golestanian2000}, but an important caveat: in that reference, a
multipole expansion is introduced at an early stage in the calculation.
This is, indeed,  adequate, in order to analyze the case of conductors when
they are separated by long distances, but it cannot be used to write an
expression of general validity. As we have shown here, such an expression
may be written in terms of the determinant of the capacitance matrix of the system. 

As shown in \cite{Golestanian2000}, in the long distance limit, the
interaction between  conductors changes drastically between the grounded
and isolated case: the former is dominated by the monopole-monopole term,
while, in the latter the leading interaction comes from the dipole-dipole
term. 

We have analyzed in some detail the behavior of $\Delta F$ in the opposite
regime to the one of \cite{Golestanian2000}, namely, at short distances.  
This is the case that should be more relevant to Casimir effect
calculations.  In that context, we note that in previous works it has been
shown that, at short distances, the corrections to PFA involve a double logarithm
behavior in the free energy, for the particular cases of a 
sphere in front of a plane \cite{Paulo}, and also for two spheres \cite{BimonteEmig2012}. 
We have shown that those double logarithmic terms come from the logarithmic
behavior of the capacitance coefficient $C_{11}$, and that their occurrence 
is not a general phenomenon, but in can nevertheless be predicted using an estimation of
the capacitance matrix based on the  PFA.   

It is worth to point out  that the short distance corrections for isolated
objects may formally be regarded as a next to NTLO correction to the PFA for
grounded ones. Indeed, we have shown this to be the case for the concrete
example of the sphere-sphere geometry.  
Note also that, in this context, when considering the corrections to the PFA
calculation of the free energy for isolated objects, one gets contributions
from two qualitatively different origins: on the one hand, one has the
terms which arise from $\Delta F$. On the other, we have the ones that
proceed from the free energy for grounded conductors. 
The derivative expansion (DE) approach \cite{denos,deothers} has been used
for the term which comes from the Dirichlet (grounded) term,  and it gives
of course the same result for either isolated and grounded conductors,
since the difference between their free energies is in $\Delta F$.

Finally, we note that there is another important difference between the
contributions to the free energy of two isolated conductors coming from the
two terms which may be identified as corresponding to grounded conductors
and to $\Delta F$. The interaction energy in the former, for
many interesting cases, can be written as a functional of the (space
dependent) vertical distance between the two surfaces. This functional becomes, in
the limit of flat and parallel conductors, extensive in their area. 
This is the starting point of the DE \cite{denos}, which for the Dirichlet
case generates a correction depending on the distance function and its
derivatives. 
The reason for this term to be extensive in the area, is that it proceeds
from the contribution of field fluctuations, which form a continuum of
degrees of freedom (to be integrated out), the number of which goes like
the area of the surfaces times the differential volume in momentum space.

The $\Delta F$ term is, on the other hand, the result of evaluating the
integral over just one (constant) mode: a single degree of freedom. 
Therefore there is no area factor in its contribution, even for flat parallel conductors.  
Even though the capacitance coefficients do depend on the areas, they
appear inside a logarithm (in spite of the fact that the PFA may be
correctly applied to calculate the capacitance coefficients). We see that
the same cannot be done to evaluate, say, the
$\Delta F$ term as the result of a single PFA (or even DE) calculation.
Indeed, as shown in \cite{denos}, the PFA is obtained as the ``effective
potential" for the corresponding functional. Namely, the ratio between the
functional and the area, in the infinite area limit, for a constant
distance between plates. And this ratio vanishes for $\Delta F$.

\section*{Acknowledgements}
This work was supported by ANPCyT, CONICET, UBA and UNCuyo.
\appendix
\section{}
\label{A}
In this Appendix we prove that the partition function that includes an integration  over the values of the surface potentials 
\begin{equation} 
{\mathcal Z}^{(i)}=\int_{-\infty}^{\infty}\left(\prod_{\alpha=1}^{N}d\phi_\alpha\right){\mathcal Z}[\{\phi_\alpha\}]\, 
\end{equation}
corresponds to an isolated conductor with fixed vanishing total charge. To this end, we introduce 
a generating functional for the mean values  $<Q_{\alpha_1}^{n_1} Q_{\alpha_2}^{n_2}....>$:
\begin{eqnarray}
{\mathcal Z}[\{\mu_\alpha\}] &=&\int_{-\infty}^{\infty}\left(\prod_{\alpha=1}^{N}d\phi_\alpha\right)
\int {\mathcal D} \phi \,
e^{- \frac{1}{2 \beta}
\int d^3{\mathbf x} \, 
(\partial_j\phi )^2 } \nonumber \\ &\times &\prod_{\alpha=1}^N\delta[\phi\vert_{S_\alpha}-\phi_\alpha] e^{-\mu_\alpha Q_\alpha} \; .
\end{eqnarray}
and prove that it does not depend on $\mu_\alpha$. The charge on each conductor reads
\begin{equation}
Q_\alpha=-\int_{S_\alpha}\vec\nabla\phi\cdot\vec{d S}\,\,\, .
\end{equation}

In order to compute the functional integral in ${\mathcal Z}[\{\mu_\alpha\}]$, we proceed as before and perform a shift in the integration variables
$\phi=\tilde\phi+\varphi$ (see Eq.(\ref{shift})). In terms of the new integration variable, $\varphi$, the charge is given by 
\begin{equation}
Q_\alpha=\tilde Q_\alpha - \int_{S_\alpha}\vec\nabla\varphi\cdot\vec{d S}\,\, ,
\end{equation}
where $\tilde Q_\alpha$ is the charge associated to the classical field $\tilde\phi$.  We obtain:
\begin{eqnarray}\label{zmu}
 {\mathcal Z}[\{\mu_\alpha
\}]&=&\int_{-\infty}^{\infty}\left(\prod_{\alpha=1}^{N}d\phi_\alpha\right)\;e^{-\frac{1}{2\beta}\sum_{\gamma\delta}C_{\gamma\delta}\phi_\gamma\phi_\delta}\nonumber \\
&\times & e^{-\sum_{\gamma\delta}\mu_\gamma C_{\gamma\delta}\phi_\delta}
 \int {\mathcal D} \varphi \, e^{- \frac{1}{2 \beta}
\int d^3{\mathbf x}  
(\partial_j\varphi )^2 } \nonumber \\ &\times & e^{\sum_\alpha \mu_\alpha \int_{S_\alpha}\vec\nabla\varphi\cdot\vec{d S}}\prod_{\alpha=1}^N\delta[\varphi\vert_{S_\alpha}]]\, \,  ,
\end{eqnarray}
which are two independent integrals. The first one (upper line in Eq.(\ref{zmu})) is an ordinary Gaussian integral. The second one (lower line in Eq.(\ref{zmu})) is a 
functional integral for a free scalar field satisfying Dirichlet boundary conditions, 
in the presence of a source $J$ defined by
\begin{equation}
 \sum_\alpha \mu_\alpha \int_{S_\alpha}\vec\nabla\varphi\cdot\vec{d S}\equiv \int d^3{\mathbf x} \;J\varphi\,\,  ,
 \end{equation} 
 so
 \begin{equation}
 J=-\sum_\alpha\mu_\alpha\int_{S_\alpha}\vec{dS_\alpha}\cdot\vec\nabla\delta({\mathbf x}-{\mathbf x_{S_\alpha}})\, \, ,
 \end{equation}
where ${\mathbf x_{S_\alpha}}$ denotes  points on the surface $S_\alpha$. 
Therefore
 \begin{equation}\label{Zmufin}
  {\mathcal Z}[\{\mu_\alpha\}] = e^{\frac{\beta}{2}\sum_{\gamma\delta}C_{\gamma\delta} \mu_\gamma\mu_\delta} e^{\frac{\beta}{2}\int d^3{\mathbf x}\int d^3{\mathbf y}J({\mathbf x})G({\mathbf x},{\mathbf y})J({\mathbf y})} \,\, ,
 \end{equation}
 where $G$ is the Green's function of the electrostatic problem 
 \begin{equation}
 \nabla^2 G({\mathbf x},{\mathbf y})=-\delta({\mathbf x}-{\mathbf y})\quad G\vert_{S_\alpha}=0\; ,
 \end{equation}
 and we omitted an overall
 constant that is independent of $\mu_\alpha$. 
 
 Using the explicit expression for the current $J$, and after integration by parts we obtain
 \begin{eqnarray}\label{JGJ}
 &&\int d^3{\mathbf x}\int d^3{\mathbf y}J({\mathbf x})G({\mathbf x},{\mathbf y})J({\mathbf y})  = \sum_{\alpha\beta}\mu_\alpha\mu_\beta\int dS_\alpha \nonumber \\ &\times &\int dS_\beta\;
 \partial_{n_\alpha} \partial_{n_\beta}G\, ,
 \end{eqnarray} 
 where we recognize the (not so well known) formal expression of the coefficients of capacitance in terms of the Green's function \cite{Uehara}
 \begin{equation}\label{CG}
 C_{\gamma\delta}=-\int dS_\gamma\int dS_\delta\;
 \partial_{n_\gamma} \partial_{n_\delta}G\; . 
 \end{equation}
 Combining Eqs.(\ref{Zmufin})-(\ref{CG}) we see that $ {\mathcal Z}[\{\mu_\alpha\}] $ does not depend on $\mu_\alpha$. Therefore, all the mean values 
 $<Q_{\alpha_1}^{n_1} Q_{\alpha_2}^{n_2}....>$ vanish.


\begin{thebibliography}{bib}
\bibitem{foot} This may not be always the case,  due to presence of the logarithm.
\end{thebibliography}
\end{document}